\documentstyle[11pt]{article}
\newcommand{\nocdot}{}
\newcommand{\noi}{\vspace{12pt}\noindent}

\newcommand{\beq}{\begin{equation}}
\newcommand{\eeq}{\end{equation}}
\newcommand{\bea}{\begin{eqnarray}}
\newcommand{\eea}{\end{eqnarray}}
\newcommand{\pa}{\partial}
\newcommand{\lpart}{\raise.3ex\hbox{$\stackrel{\leftarrow}{\partial}$}}
\newcommand{\rpart}{\raise.3ex\hbox{$\stackrel{\rightarrow}{\partial}$}}
\newcommand{\ldr}{\raise.3ex\hbox{$\stackrel{\leftarrow}{\delta^r}$}}
\newcommand{\deder}[1]{{ 
 {\stackrel{\raise.1ex\hbox{$\leftarrow$}}{\delta^r}   } 
\over {   \delta {#1}}  }}
\newcommand{\dedel}[1]{{ 
 {\stackrel{\lower.3ex \hbox{$\rightarrow$}}{\delta^l}   }
 \over {   \delta {#1}}  }}

\newcommand{\papar}[1]{{ 
 {\stackrel{\raise.1ex\hbox{$\leftarrow$}}{\partial^r}   } 
\over {   \partial {#1}}  }}
\newcommand{\papal}[1]{{ 
 {\stackrel{\lower.3ex \hbox{$\rightarrow$}}{\partial^l}   }
 \over {   \partial {#1}}  }}

\newcommand{\Hf}{{1 \over 2}}

\newcommand{\eq}[1]{{eq.~(\ref{#1})}}
\newcommand{\eqs}[2]{{eqs.\ (\ref{#1}) and (\ref{#2})}}
\newcommand{\Ref}[1]{{Ref.~\cite{#1}}}
\newcommand{\mb}[1]{{\mbox{${#1}$}}}
\begin{document}
\thispagestyle{empty}
\vspace{3cm}
\title{\Large{\bf Superfield Formulation of the Phase Space Path Integral 
}}
\vspace{2cm}
\author{{\sc I.A.~Batalin}\\
Lebedev Physics Institute\\
53 Leninisky Prospect\\Moscow 117924\\Russia\\
~\\~{\sc K.~Bering}\\
Institute for Fundamental Theory \\ 
Department of Physics \\
University of Florida \\
Gainesville, FL, 32611, USA\\~\\and\\~\\
{\sc P.H.~Damgaard}\\The Niels Bohr Institute\\Blegdamsvej 17\\
DK-2100 Copenhagen\\Denmark}
\maketitle
\begin{abstract}
We give a superfield formulation of the path integral on an arbitrary 
curved phase space, with or without first class constraints. Canonical 
tranformations and BRST transformations enter in a unified manner. 
The superpartners of the original phase space variables precisely conspire 
to produce the correct path integral measure, as Pfaffian ghosts. When 
extended to the case of second-class constraints, the correct path integral 
measure is again reproduced after integrating over the superpartners. These
results suggest that the superfield formulation is of first-principle
nature.
\end{abstract}
\vspace{1.8cm}
\begin{flushleft}
NBI-HE-98-34\\ UFIFT-HEP-98-40  \\hep-th/9810235
\end{flushleft}

\vfill
\newpage

\noi
{\em 1.\ Introduction}.
In a recent paper \cite{super}, we have shown that an arbitrary Hamiltonian 
quantum field theory can be given a superfield formulation.
Although the formalism of \Ref{super} and the constructions explained below  
can be formulated in operator language, we shall here 
focus on the path integral formalism.
The needed superspace is two-dimensional, consisting of time $t$ and a new
Grassmann-odd superpartner, which we denote by $\theta$. All original 
phase space coordinates \mb{z^{A}_{0}(t)} are then treated as zero-components 
of super phase space coordinates
\beq
z^{A}(t,\theta) ~=~ z^{A}_{0}(t) + \theta z^{A}_{1}(t) ~.
\label{z}
\eeq
In particular, \mb{z^{A}(t,\theta)} has the same statistics as 
\mb{z^{A}_{0}(t)}, which we denote by \mb{\epsilon_{A}}. 
One essential ingredient of \Ref{super} was the introduction of a superspace 
derivative 
\beq
D ~\equiv~ \frac{d}{d\theta} + \theta \frac{d}{dt} ~,
\label{D}
\eeq
which acts like a ``square root'' of the time derivative:
\beq
D^2 ~=~ \frac{d}{dt} ~.
\eeq
The superspace extends in an obvious manner to a \mb{(d+1)}-dimensional 
superspace of coordinates \mb{(x^{\mu},\theta)} when considered in the 
context of a Lorentz invariant quantum field theory 
in $d$ dimensions, but we shall here restrict ourselves to the 
finite-dimensional case of $2N$ phase space variables.

\noi
The purpose of the present paper is to demonstrate that the superspace 
formalism developed in \cite{super} reaches one step deeper than could 
have been anticipated. 
By considering here the extension to a phase space with a non-constant
symplectic metric,
we shall show that the required superspace generalization of the phase 
space path integral \cite{BF0,BF1} leads, after integrating out the 
fermionic coordinate $\theta$, to precisely the correct path integral measure. 
This is a quite non-trivial fact, completely independent of whether we 
consider a system with (first class) constraints or not. 
Moreover, when considered in the presence of second-class constraints
it turns out that our formalism also here directly yields all required 
factors in the path integral. 

\noi
{\em 2.\ Symplectic Structure}.
In addition to \eqs{z}{D}, the few ingredients we need are as follows. 
Define a graded Poisson bracket by
\beq
\{F,G\} 
~\equiv~ F\lpart_{A}\omega^{AB}\rpart_{B} G ~,\label{pb}
\eeq
for functions \mb{F=F(z(t,\theta))}, \mb{G=G(z(t,\theta))}.
Here the (non-degenerate) symplectic metric, 
\beq
\omega^{AB}~=~\omega^{AB}\left(z(t,\theta)\right) ~=~ 
\{z^{A}(t,\theta),z^{B}(t,\theta)\} ~,
\eeq
is allowed to depend on \mb{z^{A}(t,\theta)}. 
We will in what follows suppress some of the arguments to
make the formulas more readable. For precise details we refer to the 
original paper \cite{super}. 
The symmetry properties are as follows:
\beq
\omega^{BA} ~=~ - (-1)^{\epsilon_{A}\epsilon_{B}}\omega^{AB} ~~,~~~~~~
\epsilon(\omega^{AB}) = \epsilon_{A} + \epsilon_{B} ~,
\eeq
which ensures
\beq
\{F,G\} ~=~ -(-1)^{\epsilon(F)\epsilon(G)}\{G,F\} ~.
\eeq
Similarly, the super Jacobi identity
\beq
\{\{F,G\},H\}(-1)^{\epsilon(F)\epsilon(H)}~ + 
~{\mbox{\rm cyclic perm.(F,G,H)}} ~=~ 0 ~,
\eeq
is satisfied if
\beq
\omega^{AD}\partial_{D} \omega^{BC}(-1)^{\epsilon_{A}\epsilon_{C}}~ + 
~{\mbox{\rm cyclic perm.(A,B,C)}} ~=~ 0 ~.
\eeq
As usual, we define an inverse symplectic metric \mb{\omega_{AB}} by
\mb{\omega^{AB}\omega_{BC} = \delta^{A}{}_{C}}. Its symmetry properties are
quite different:
\beq
\omega_{BA} ~=~ (-1)^{(\epsilon_{A}+1)(\epsilon_{B}+1)}\omega_{AB} ~~,~~~~~~
\epsilon\left(\omega_{AB}\right) = \epsilon_A + \epsilon_B ~.
\eeq
Crucial in this context is that the Jacobi identity turns into a
{\em closedness} relation
\beq
\partial_{C}\omega_{AB}(-1)^{(\epsilon_{C}+1)\epsilon_{B}}~ + 
~{\mbox{\rm cyclic perm.(A,B,C)}} ~=~ 0 ~,
\eeq
which implies that locally we can represent 
$\omega_{AB}$ in terms of a symplectic potential $V_{A}$:
\beq
\omega_{AB} ~=~ \left(\partial_{A} V_{B} 
 -(-1)^{\epsilon_{A}\epsilon_{B}}\partial_{B} V_{A}\right)
(-1)^{\epsilon_{B}} ~.
\label{rot}
\eeq
Our primary aim is not to elaborate on global issues. We shall for simplicity
assume that the phase space is simply connected and that there exists a 
globally defined symplectic potential.

\noi
{\em 3.\ Super Hamiltonian}.
Let there now be given a Grassmann-odd BRST generator 
\mb{\Omega = \Omega(z(t,\theta))} and an
Hamiltonian \mb{H = H(z(t,\theta))} with the properties
\cite{BFV}
\beq
\{\Omega,\Omega\} ~=~ 0
 ~~~~~~~~{\mbox{\rm and}}~~~~~~~~~ \{H,\Omega\}~=~ 0~.
\label{homegapb}
\eeq
We combine these two fundamental objects into a Grassmann-odd superfield $Q$:
\beq
Q(z(t,\theta),\theta) ~\equiv~ \Omega(z(t,\theta)) + 
\theta H(z(t,\theta)) ~.
\label{qdef}
\eeq
It is nilpotent in terms of the Poisson bracket, by virtue of eq. 
(\ref{homegapb}):
\beq
\{Q,Q\} ~=~  0 ~.\label{qpb}
\eeq
This nilpotency condition is preserved under super canonical transformations
\beq
Q ~\mapsto~ Q_{\Psi} ~\equiv~
e^{{\mbox{\rm ad}} \Psi}Q ~,
\label{qpsi}
\eeq
which infinitesimally are generated by the adjoint action
\beq
{\mbox{\rm ad}} ~\Psi  \equiv \{\Psi,~\cdot~\} ~. 
\eeq
Here $\Psi$ is a superfield,
\beq
\Psi\left(z(t,\theta),\theta\right) ~=~ \Psi_0\left(z(t,\theta)\right)
+ \theta\Psi_1\left(z(t,\theta)\right)~,
\eeq
which plays the r\^{o}le of a generalized gauge-fixing fermion. 
More precisely, $\Psi$ itself is Grassmann-{\em even}, and it is the 
$1$-component $\Psi_{1}$ which directly corresponds to the gauge-fixing 
fermion. Instead, the bosonic zero-component $\Psi_{0}$ is a generator of 
ordinary canonical transformations \cite{super}.

\noi
{\em 4.\ The Action}.
The classical equations of motion are taken to be 
\beq
Dz^{A}
~=~ -\{Q_{\Psi},z^{A}\} ~.
\label{shpb}
\eeq
As was shown in ref. \cite{super}, these reduce to the standard equations
of motion in the original phase space variables $z_0^A(t)$. An action
which yields these equations of motion is
\beq
S[z] ~=~ \int_{t_{i}}^{t_{f}}   \! dt~ d\theta \left[
 z^{A} ~\bar{\omega}_{AB}~
Dz^{B}(-1)^{\epsilon_{B}} 
- Q_{\Psi}\right] 
\label{susyaction}
\eeq
where 
\beq
\bar{\omega}_{AB} ~\equiv~ \left(z^{C}\partial_{C} + 2\right)^{-1}\omega_{AB}
~=~ \int_{0}^{1}\! ~\omega_{AB}(\alpha z)~\alpha d\alpha ~.
\label{omegabar}
\eeq
Note that we regain the well-known kinetic term in the case of a constant
$\omega_{AB}$ for which \mb{\bar{\omega}_{AB}=\Hf \omega_{AB}}.
We may rewrite the action as
\beq
S[z] ~=~ \int_{t_{i}}^{t_{f}} \! dt~ d\theta \left[
 V_A~Dz^{A} - Q_{\Psi}\right] - 
\left[W(z(t,0))\right]^{t_{f}}_{t_{i}}  ~,
\label{susyaction1}
\eeq
where the boundary term is given by
\beq
  W(z)~\equiv~ z^A\bar{V}_{A}~,
\eeq
and
\beq
 \bar{V}_{A} ~\equiv~ \left(z^{C}\partial_{C} + 1\right)^{-1}V_{A} 
~=~ \int_{0}^{1}\! V_{A}(\alpha z)~ d\alpha ~.
\eeq
{}From \eqs{rot}{omegabar} it follows that
\beq
\bar{\omega}_{AB} ~=~ \left(\partial_{A}\bar{V}_{B} - 
(-1)^{\epsilon_{A}\epsilon_{B}}\partial_{B}\bar{V}_{A}\right)
(-1)^{\epsilon_{B}} ~.
\eeq

\noi
{\em 5.\ Partition Function}.
We therefore take the action \eq{susyaction}, or equivalently 
\eq{susyaction1}, as the correct candidate to be
exponentiated, and integrated over in the superfield path integral:
\beq
{\cal Z} ~=~ \int [dz] \  \exp\left[\frac{i}{\hbar}S[z]\right] ~.
\label{susypathint}
\eeq
Note that this path integral contains no additional measure factors.
This is not needed because the measure \mb{[dz]} remarkably transforms as
a scalar under general coordinate transformations, due to the balance between
bosonic and fermionic degrees of freedom in the superfield formulation.
In this case, on a curved phase space manifold, a crucial test of the 
present formalism is to see if we recover the correct path integral measure 
after integrating out the fermionic $\theta$-coordinate. The calculation
is straightforward, and results in $S = S_0 + S_1$ with
\bea
S_{0} &=& \int_{t_{i}}^{t_{f}} \! dt~ \left[ \dot{z}_{0}^{A}V_A(z_{0})
- H_{\Psi}(z_{0})  \right] -
\left[W(z(t,0))\right]^{t_{f}}_{t_{i}}  \cr
S_{1} &=& \int_{t_{i}}^{t_{f}} \! dt~ 
\left[\Hf z_{1}^{A} \omega_{AB}(z_{0}) z_{1}^{B}(-1)^{B} 
- z_{1}^{A} \pa_{A}\Omega_{\Psi}(z_{0})\right] ~. 
\eea
Here $H_{\Psi}$ and $\Omega_{\Psi}$ are defined according to 
\eq{qdef} and \eq{qpsi}.
After a gaussian integration over the superpartner \mb{z_{1}^{A}}, and use 
of the nilpotency relation \mb{\{\Omega_{\Psi},\Omega_{\Psi} \}=0}, 
one arrives at the standard form
\beq
{\cal Z} ~=~ \int [dz_{0}]~ {\rm Pf} (\omega_{\nocdot \nocdot})~  
\exp\left[\frac{i}{\hbar}S_{0}[z_{0}]\right] ~,
\label{comppathint}
\eeq
where the Pfaffian of an arbitrary even supermatrix is given by 
${\mbox{\rm Pf}}(M) = ({\mbox{\rm Ber}}(M))^{1/2}$.

\noi
{\em 6.\ Second Class Constraints}. To test yet again how fundamental the 
present superfield formulation is, let us now consider the case of
$2n$ second class constraints \mb{\Phi_{\alpha} = \Phi_{\alpha}(z(t,\theta))}
of Grassmann parity
\mb{\epsilon_{\alpha}}. To impose such constraints in the path integral,
we introduce an auxiliary superfield
\beq
\lambda^{\alpha}(\theta)~=~\lambda^{\alpha}_{0}+ \theta \lambda^{\alpha}_{1} 
\eeq
of Grassmann parity \mb{\epsilon_{\alpha}+1}, and consider
the partition function 
\beq
{\cal Z} ~=~ \int [dz][d\lambda] \  
\exp\left[\frac{i}{\hbar}\left(S[z] +  \int_{t_{i}}^{t_{f}}   \!  
\!dt~d\theta~ \Phi_{\alpha}\lambda^{\alpha} \right) \right] ~.
\label{susypathint2nd}
\eeq
Note again the absence of any non-trivial measure factors.
Let us show that this superfield partition function completely 
reproduces the stardard version of the partition function with second 
class constraints. 
The crucial property of second-class constraints is that the matrix
\beq
    \omega_{\alpha\beta}~=~ \{ \Phi_{\alpha}, \Phi_{\beta} \}
~=~ -(-1)^{\epsilon_{\alpha}\epsilon_{\beta}} \omega_{\beta\alpha}
\eeq 
is invertible. Let us denote the inverse matrix 
\beq
    \omega^{\alpha\beta}
~=~ (-1)^{(\epsilon_{\alpha}+1)(\epsilon_{\beta}+1)} \omega^{\beta\alpha}~.
\eeq 
According to the standard Dirac procedure, the Poisson bracket should be
replaced by the Dirac bracket:
\beq
   \{F,G\}_{D}~=~ \{F,G\} 
- \{F, \Phi_{\alpha}\}\omega^{\alpha\beta}\{ \Phi_{\beta}, G\}~.
\eeq

\noi
Let us now trace the additional terms in the path integral due to the
second-class constraints. We do this as before by integrating over the
fermionic coordinate $\theta$. The result is as follows. First, the zero 
component part \mb{S_{0}} of the action picks up a delta 
function term that precisely enforces the second-class constraints in the 
original phase space variables:
\beq
    S_{0} ~\to~ S_{0}+
\int_{t_{i}}^{t_{f}} \!  dt~ 
\lambda_1^{\alpha}\Phi_{\alpha}(z_{0}) ~.
\eeq
In the \mb{S_{1}}-part of the action, integration over $\theta$ effectively
just corresponds to replacing  
\mb{\Omega_{\Psi}(z_{0})} by  
$$
 \Omega_{\Psi}(z_{0}) 
- \Phi_{\alpha}(z_{0})  \lambda^{\alpha}_{0} ~.
$$
Therefore the gaussian integration over the superpartner \mb{z_{1}^{A}},
besides yielding the correct Pfaffian 
\mb{{\rm Pf} (\omega_{\nocdot \nocdot})} as before, also produces a term
\beq
 \Hf \left\{  \Omega_{\Psi}(z_{0}) 
- \Phi_{\alpha}(z_{0})  \lambda^{\alpha}_{0}~,~
\Omega_{\Psi}(z_{0}) 
- \Phi_{\beta}(z_{0})  \lambda^{\beta}_{0} \right\}
\eeq
in the action by completing the square.\footnote{Without second-class
constraints this term was just $\Hf\{\Omega_{\Psi},\Omega_{\Psi}\}$, which
in that case would vanish on account of the nilpotency condition
\eq{homegapb}.} If we next perform also the
gaussian integration over the zero component \mb{\lambda^{\alpha}_{0}},
we get \mb{{\rm Pf} ( \{ \Phi_{\nocdot},\Phi_{\nocdot} \})}. 
The rest of the action conspires to yield \cite{BFV}
\beq
\Hf\{\Omega_{\Psi},\Omega_{\Psi}\} - \Hf\{\Omega_{\Psi},\Phi_{\alpha}\}
\omega^{\alpha\beta}\{\Phi_{\beta},\Omega_{\Psi}\} ~=~ 
    \Hf\{\Omega_{\Psi},\Omega_{\Psi}\}_{D}~=~ 0~.
\eeq 
Therefore we quite remarkably arrive at just the standard form of the 
partition function \cite{sen}:
\beq
{\cal Z} ~=~ \int [dz_{0}]~ {\rm Pf} (\omega_{\nocdot \nocdot})~  
\exp\left[\frac{i}{\hbar}S_{0}[z_{0}]\right] \delta(\Phi_{\nocdot})  
~{\rm Pf} ( \{ \Phi_{\nocdot},\Phi_{\nocdot} \})   ~.
\label{comppathint2nd}
\eeq

\noi
{\em 7.\ Conclusions}.
The superfield formulation introduced in \cite{super} thus in a very
precise and non-trivial manner encodes all the information required
for Hamiltonian path integral quantization for systems with or without
any combination of first and second class constraints, on an arbitrary
curved phase space. In view of this, we propose to consider our superfield
formalism as a first principle on which to base quantization. An
operatorial formulation of precisely the same superfield formulation
also exists, with or without first and second class constraints,
and with possibly non-constant symplectic $\omega_{AB}$.

\vspace{1cm}
\noindent
{\sc Acknowledgement:}~I.A.B.\ and K.B.\ would like to thank the Niels Bohr 
Institute for the warm hospitality extended to them there.
The work of I.A.B.\ and P.H.D.\ is partially supported by grant INTAS-RFBR
95-0829. I.A.B.\  also acknowledges the funding by grants 
INTAS 96-0308,
RFBR 96-01-00482 and RFBR 96-02-17314. The work of K.B.\ is supported by DoE 
grant DE-FG02-97ER41029 and Nordita.


\begin{thebibliography}{999}
\bibitem{super}I.A.~Batalin, K.~Bering and P.H.~Damgaard, Nucl.~Phys.\
{\bf B515} (1998) 455.
\bibitem{BF0}I.A.~Batalin and E.S.~Fradkin, Mod.~Phys.~Lett.\ {\bf A4}
(1989) 1001.
\bibitem{BF1}I.A.~Batalin and E.S. Fradkin, Nucl. Phys. {\bf B326}
(1989) 701.
\bibitem{BFV}E.S.~Fradkin and G.A.~Vilkovisky, Phys.~Lett.\ {\bf B55}
(1975) 224.\newline
I.A.~Batalin and G.A.~Vilkovisky, Phys.~Lett. \ {\bf B69} (1977) 309. \newline
E.S.~Fradkin and E.S.~Fradkina, Phys.~Lett. \ {\bf B72} (1978) 343. \newline
I.A.~Batalin and E.S.~Fradkin, Phys.~Lett. \ {\bf B122} (1983) 157.
\bibitem{sen}E.S. Fradkin, {\em in Proc. X Winter School of Theo. Phys.},
Karpacs (1973) No. 207, pp. 93.\newline
P. Senjanovic, Ann. Phys. {\bf 100} (1976) 227.




\end{thebibliography}
\end{document}